\documentclass[conference]{IEEEtran}

\usepackage[dvipsnames]{xcolor}
\usepackage{cite}
\usepackage{graphicx}
\usepackage{cite}
\usepackage{graphicx}
\usepackage{graphicx}
\usepackage{float}
\usepackage[spaces,hyphens]{url}
\usepackage[colorlinks,allcolors=blue]{hyperref} 

\usepackage{graphicx}
\usepackage{subcaption}
\usepackage[margin=1in]{geometry}
\usepackage{tikz}
\usepackage[utf8]{inputenc} %
\usepackage[T1]{fontenc}  
\usepackage[english]{babel}
\usetikzlibrary{positioning, arrows.meta} \usepackage{amsmath}
\ifCLASSINFOpdf
\else
\fi

\begin{document}

\title{Fuzzy Intelligent System for Student Software Project Evaluation}

\author{\IEEEauthorblockN{Anna Ogorodova \IEEEauthorrefmark{1},
Pakizar Shamoi\IEEEauthorrefmark{2}, Aron Karatayev\IEEEauthorrefmark{3}}
\IEEEauthorblockA{School of Information Technology and Engineering \\
Kazakh-British Technical University\\
Almaty, Kazakhstan\\
Email: \IEEEauthorrefmark{1}an\_ogorodova@kbtu.kz,
\IEEEauthorrefmark{2}p.shamoi@kbtu.kz,
\IEEEauthorrefmark{3}ar\_karatayev@kbtu.kz
}
}

\maketitle

\IEEEpeerreviewmaketitle

\begin{abstract}



Developing software projects allows students to put knowledge into practice and gain teamwork skills. However, assessing student performance in project-oriented courses poses significant challenges, particularly as the size of classes increases. The current paper introduces a fuzzy intelligent system designed to evaluate academic software projects using object-oriented programming and design course as an example. To establish evaluation criteria, we first conducted a survey of students (n=31) and faculty (n=3) to identify key parameters and their applicable ranges. The selected criteria—clean code, use of inheritance, and functionality—were selected as essential for assessing the quality of academic software projects. These criteria were then represented as fuzzy variables with corresponding fuzzy sets. Collaborating with three experts, including one professor and two course instructors, we defined a set of fuzzy rules for a fuzzy inference system. This system processes the input criteria to produce a quantifiable measure of project success. The system demonstrated promising results in automating the evaluation of projects. Our approach standardizes project evaluations and helps to reduce the subjective bias in manual grading.

\end{abstract}

\section{Introduction} 
Academic software projects are essential for information technology students to gain hands-on experience, real-world applications, teamwork, and portfolio building. Educational institutions, especially technical universities, often offer courses that teach programming. These courses require students to undertake projects - develop code to solve specific problems.  While courses like Algorithms and Data Structures or Fundamentals of Programming might use automated assessments through input and output files, evaluating student performance in project-oriented courses like Object-oriented programming is more complex due to the diverse nature of project work.
Therefore, given the importance of software projects in an academic environment evaluation and feedback from mentor-teachers is increasingly taking a key position in education \cite{Rafiq2022}.

In general, artificial intelligence (AI) has a lot of potential applications in education, particularly in the areas of tutoring, assessment, and personalization \cite{app11125467}. Another important feature of AI in education is the ability to grade students automatically \cite{app11125467}. Providing students with timely and accurate feedback through qualitative assessment improves their learning in a higher education setting \cite{Hooda2022}.

\begin{figure*}[htbp]
\centering
\resizebox{\textwidth}{!}{%
\begin{tikzpicture}[node distance=0.4cm and 2cm, auto, scale=0.8, transform shape]
    \tikzset{>=Latex}
    
    \tikzset{block/.style={rectangle, draw, text width=3.5cm, align=center, minimum height=1cm, font=\small}}
    \node[block] (A) {Increasing number of students};
    \node[block, below=of A] (B) {Instructor workload increase};
    \node[block, below=of B] (C) {Less complete project analysis};
    \node[block, below=of C] (D) {Missed errors, incomplete feedback};
    \node[block, right=of B] (E) {Variation in student knowledge levels};
    \node[block, below=of E] (F) {Difficulty in providing individualized feedback};
    \node[block, right=of E] (G) {Limited resources};
    \node[block, below=of G] (H) {Decreased objectivity of grades};
    \node[block, below=of H] (I) {Difficulty in analyzing complex projects};
    \node[block, right=of G] (J) {Subjectivity in evaluation criteria};
    \node[block, below=of J] (K) {Differences in interpretation by evaluators};
    \node[block, below=of K] (L) {Confusion and uncertainty for students};
    \node[block, below=of A, yshift=-4.25cm] (M) {Difficulty in student understanding of feedback};
    \node[block, below=of M] (N) {Decreased motivation and learning};

    \draw[->] (A) -- (B);
    \draw[->] (B) -- (C);
    \draw[->] (C) -- (D);
    \draw[->] (D) -- (M);
    \draw[->] (M) -- (N);
    \draw[->] (B) -- (E);
    \draw[->] (E) -- (F);
    \draw[->] (E) -- (G);
    \draw[->] (G) -- (H);
    \draw[->] (H) -- (I);
    \draw[->] (G) -- (J);
    \draw[->] (J) -- (K);
    \draw[->] (K) -- (L);
\end{tikzpicture}
}
\caption{Flowchart depicting the challenges of evaluating academic software projects}
\label{fig:flowchart}
\end{figure*}
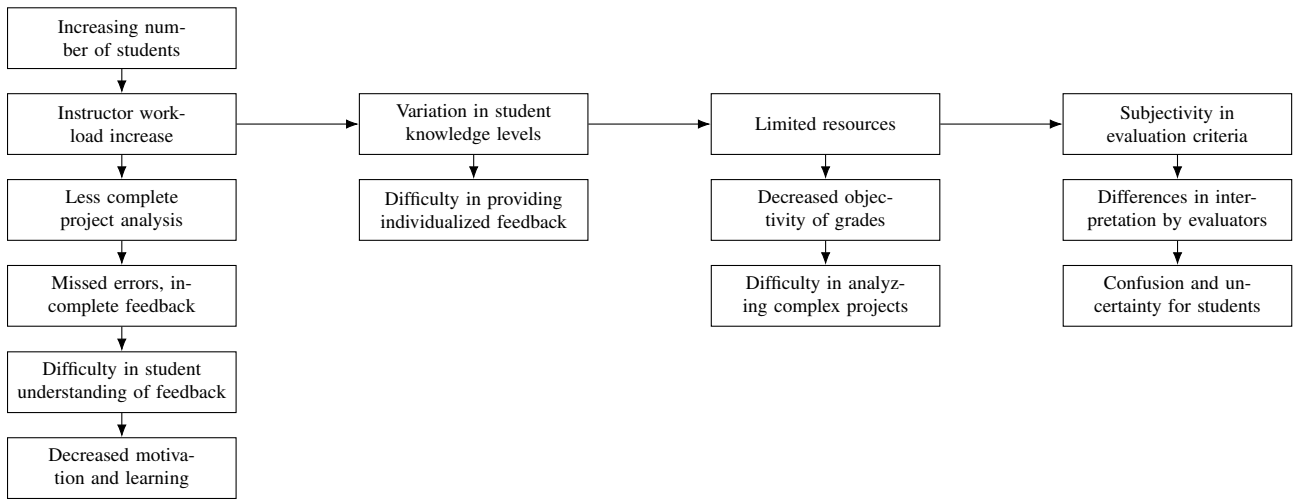

Assessing student performance in project-oriented courses is a complex task that demands careful attention. As the number of students in technical programs increases, the instructor's workload in checking student software projects also increases. This is due to time constraints, varying levels of student knowledge, limited resources, and the complexity of the projects (see Fig. \ref{fig:flowchart}). The instructor's overload can lead to less complete project analysis and missed errors, subjective grading, or incomplete feedback. 





The next problem that may arise when evaluating academic projects is limited resources. As the number of students increases, the instructor may need more resources to validate software projects, such as grading software or more teaching assistants. Assuming each project takes 30 minutes, an instructor would need about 150 hours to review 300 projects. The teacher may also need additional time to provide feedback, assign grades, and communicate with students about their projects.  In addition, with an increased number of students the instructor will likely encounter a broader range of comprehension and skill levels \cite{Yilmaz2017}, which makes it difficult to provide individualized feedback appropriate to each student's level of understanding and skill.  Delays in feedback can lead to decreased engagement, motivation, and increased anxiety among students, which can significantly hinder their learning experience \cite{Galevska2019}. 




The problem of software project estimation is a complex issue requiring careful consideration of many factors \cite{Zhou2001}. Software projects may sometimes need clear objectives that can be easily measured. In an academic context, success may be defined differently, depending on the goals of the project. For example, a program project may be evaluated on its technical merit, the student group's work, communication skills, and knowledge of theoretical material. Each of these factors requires its own set of evaluation criteria, and they can be challenging to measure objectively. Evaluating a software project can be subjective, as instructors may have different opinions about what constitutes success or failure \cite{Amelia2019}. For example, students may consider a project a success if it meets technical requirements. In contrast, faculty may consider it a success if all team members are equally involved in its development. Particular attention should be paid to another significant problem, evaluation uncertainty, which arises when more than one expert or evaluator is involved in the evaluation process. Differences in understanding assessment criteria and subjective views can lead to differences in final grades, creating confusion and uncertainty for students. It is crucial to develop evaluation criteria consistent with the project's goals and to consider the context of the academic environment. 



\begin{figure}[h]
  \includegraphics[width=0.5\textwidth]{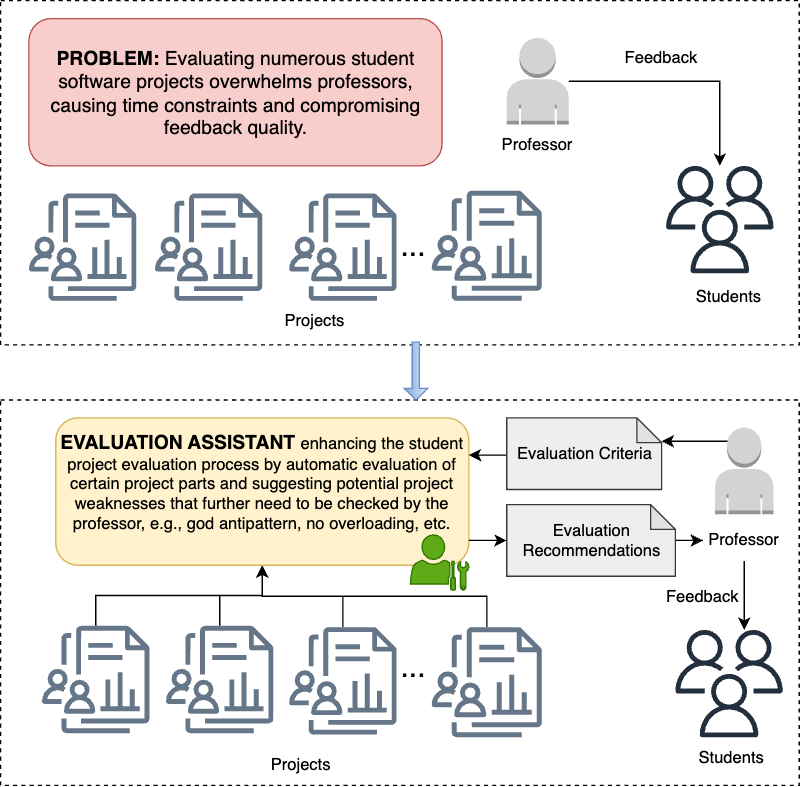}
  \caption{Idea of evaluation assistant}
  \label{evass}
\end{figure}


In this context, developing a fuzzy intelligent system for evaluating student software projects can address these problems. Such a system can automate the evaluation process, making it more objective, consistent, and transparent, a kind of evaluation assistant system (see Fig.\ref{evass}). Developing software is by its nature imprecise \cite{multi}. The motivation for using Fuzzy Logic in this problem is that it can handle ambiguity and uncertainty, incorporate expert knowledge, and combine multiple criteria into an assessment \cite{karatayev2024fuzzy}. In addition, fuzzy logic will allow the system to better adapt to the diversity of student projects.


This paper introduces an evaluation model for software development projects that utilize fuzzy logic to address the uncertainty resulting from human subjective perception  during decision-making. The main contributions of this study are:

\begin{itemize}
    \item Identifying the critical criteria for evaluating academic software projects based on survey and real academic projects
    \item Development of an intelligent system for evaluating software projects, its assessment done by experts.
\end{itemize}

The structure of the paper is as follows. Section I is this Introduction. Section II presents a thorough analysis of previous studies of academic project evaluation. Section III describes research methods, including an explanation of the fuzzy sets and logic that serve as the basis system. The section also covers the data collection procedure and the surveys used to identify the evaluation criteria. Results are presented in section IV. The study's conclusions and recommendations for future improvements are presented in Section V.

\section{Related Work}
The current section provides a review of the existing literature on AI methods (specifically, fuzzy sets and logic) for evaluating software projects.

Developing assessment strategies and techniques that can facilitate learning and teaching effectively has been the subject of extensive research \cite{Hooda2022}. Specifically, AI has been widely applied in the field of education \cite{ieeeaccess}.

Several works propose using fuzzy logic theory to evaluate students' performance \cite{fuzzyLogic, Saturday2019}. 

The authors used a criterion-based approach to evaluate student projects based on experiments. Students' work was graded according to a list of pre-agreed grading criteria developed by instructors in collaboration with students \cite{Ajol2020}. The authors allow users to modify the main and sub-criteria and their weights in decision-making systems according to their evaluation priorities. An objective multi-criteria decision-making system for evaluating the effectiveness and problem-oriented concepts in education has been proposed  \cite{Fuzzy2012}. A survey questionnaire consisting of open-ended questions was also conducted in some studies to see the effectiveness of the study and get feedback. 

Another study used fuzzy sets to determine the evaluation criteria and their corresponding weights. The matched criteria are then used to assess student learning outcomes \cite{Zhou2000}. The authors propose various criteria, such as acceptability, number of program classes, test coverage, and effectiveness, to help instructors evaluate program projects according to the criteria, given the strengths and limitations of the preferred project evaluation model, and to help project evaluators understand the logic behind different approaches to project evaluation \cite{Haass2020}. The problem of assessing students' academic performance using the fuzzy logic model has been considered in \cite{Barlybayev2016}. Their research was based on assessments such as grades in lectures, practical classes, students' independent work, and laboratory work as criteria for academic performance. The other study introduced the fuzzy assessment system for distance learning that analyzes student performance, behavior, and exams \cite{new1}. 
Specification of teaching activity using fuzzy logic was introduced in \cite{Putera2015}.

Several studies considered the idea of automatic grading of students' projects \cite{mod1}, \cite{mod2}. The method for automatically evaluating and grading student UML diagrams was recently proposed \cite{mod1}. It employs a Java-based algorithm that processes the instructor's and the student's solution diagrams, subsequently generating the student's scores while detecting mistakes. The other study utilized the hybrid approach, fuzzy logic, and hierarchical linear regression to evaluate students’ performance \cite{mod2}.



Evaluating academic software can be complex and multifaceted. Intelligent systems may need help to handle such a complex evaluation, especially if multiple criteria must be considered. Some aspects of academic software evaluation, such as user experience and interface design, are subjective. It can be difficult for an intelligent system to handle subjective evaluations because they vary from user to user. The fuzzy approach has also been used in an assessment model that builds upon the VIKOR compromise ranking method and uses the fuzzy multi-criteria decision-making (MCDM) approach to gauge the success of software development projects  \cite{multi}. Another study focused on applying the association rules for project evaluation \cite{Grzeszczyk2013}.




A recent work \cite{Hooda2022} provides a comparative analysis of how AI can improve student learning outcomes through assessment and feedback procedures. The study provides an overview of the most popular AI and ML algorithms for student success. According to the results, I-FCN outperformed other methods (ANN, XG Boost, SVM, Random Forest, and Decision Trees). Fuzzy Logic was not used in this analysis. More recent work by \cite{IZBASSAR2024779} used a hybrid approach  (ML models and fuzzy sets)  to evaluate students’ readiness for post-graduation challenges using surveys.

As we see, neural networks, deep learning, random forest, logistic regression, multilayer perceptron, naive Bayes, support vector machines, decision trees, and fuzzy methods have all been used in studies for the assessment and evaluation of student performance evaluation in the literature. However, most studies use subjectively defined criteria, and limited works provide ways to customize these methods to specific courses and experts. Additionally, despite the numerous research studies on student performance evaluation, only a few works focused on engineering project evaluation.

\section{Methods}
The creation of an intelligent system is divided into several stages, including data collection, definition of evaluation criteria, system design, and development. Data collection includes gathering relevant information from academic software projects via surveys of students and experts (teachers). Questionnaire responses will be collected and analyzed to determine key evaluation factors and their importance. The system design phase focuses on defining the architecture and functionality of the intelligent system. The development phase involves writing code and building the intelligent system. 

\subsection{Fuzzy Sets and Logic}
Fuzzy set theory will be used as the basis for the evaluation model. Lotfi Zadeh introduced fuzzy sets in the 1960s to represent uncertainty and fuzziness in natural language expressions \cite{ZADEH1965338}. Fuzzy sets are used in various applications, such as decision-making, control systems, pattern recognition, and artificial intelligence. Fuzzy logic allows the representation of imprecise and uncertain information often found in software project evaluation. Fuzzy sets will be used to define evaluation criteria and linguistic variables.

\subsubsection{Membership Functions and Fuzzy Sets}
Fuzzy sets, first introduced by Zadeh \cite{ZADEH1965338}, allow degrees of membership, which are indicated 
with a number between 0 and 1. So, in contrast to the pair of numbers {\{}0,1{\}} in Boolean logic, we move to all the numbers in a range [0,1]. This is called a 
\textit{membership function} (MF) and is denoted as $\mu _{A}$(x) and, in this way, can denote fuzzy sets.
MFs are mathematical techniques for modeling the meaning of symbols by indicating flexible membership to a set. We can use it to represent uncertain concepts like age, performance, building height, etc. Therefore, MF's essential function is to convert a crisp value to a membership level in a fuzzy set. 

The shape of the membership function reflects the degree of fuzziness or uncertainty of the set. We use triangular and trapezoidal MF in this study, which are represented as follows.
The triangular membership function is defined by three parameters \( a \), \( b \), and \( c \), where \( a \leq b \leq c \). It is described by the following piecewise function:

\begin{tikzpicture}
\draw[thick,->] (0,0) -- (6,0) node[anchor=north west] {$x$};
\draw[thick,->] (0,0) -- (0,5) node[anchor=south east] {$\mu_A(x)$};

\draw[thick] (1,0) node[below] {$a$} -- (3,4) node[above] {$b$} -- (5,0) node[below] {$c$} -- cycle;

\draw[dotted] (3,0) node[below] {$b$} -- (3,4);
\end{tikzpicture}

\[
\mu_{\text{triangular}}(x; a, b, c) = 
\begin{cases} 
\frac{x-a}{b-a} & \text{if } a \leq x < b, \\
\frac{c-x}{c-b} & \text{if } b \leq x < c, \\
0 & \text{otherwise.}
\end{cases}
\]
This function increases linearly from 0 at \( x = a \) to 1 at \( x = b \) and decreases back to 0 at \( x = c \).

The trapezoidal membership function is defined by four parameters \( a \), \( b \), \( c \), and \( d \), where \( a \leq b \leq c \leq d \). It is described by the following piecewise function:

\begin{tikzpicture}
\draw[thick,->] (0,0) -- (7,0) node[anchor=north west] {$x$};
\draw[thick,->] (0,0) -- (0,5) node[anchor=south east] {$\mu_B(x)$};

\draw[thick] (1,0) node[below] {$a$} -- (2,4) node[above left] {$b$} -- (4,4) node[above right] {$c$} -- (5,0) node[below] {$d$};

\draw[dotted] (2,0) node[below] {$b$} -- (2,4);
\draw[dotted] (4,0) node[below] {$c$} -- (4,4);
\end{tikzpicture}
\[
\mu_{\text{trapezoidal}}(x; a, b, c, d) = 
\begin{cases} 
\frac{x-a}{b-a} & \text{if } a \leq x < b, \\
1 & \text{if } b \leq x \leq c, \\
\frac{d-x}{d-c} & \text{if } c < x \leq d, \\
0 & \text{otherwise.}
\end{cases}
\]
This function increases linearly from 0 at \( x = a \) to 1 at \( x = b \), stays constant at 1 between \( x = b \) and \( x = c \), and decreases back to 0 at \( x = d \).


Representing linguistic terms and hedges, or linguistic expressions that modify other expressions, is a significant component of the fuzzy set theory framework \cite{SHAMOI2020217}. A fuzzy set typically represents a linguistic term, and an operation that changes one fuzzy set into another represents a linguistic modifier or hedge. 

\subsubsection{Linguistic Variables}
According to Zadeh \cite{Zadeh1975}, ``By a linguistic variable we\textbf{\textit{ 
}}mean a variable whose values are not numbers but words or sentences in a 
natural or artificial language''. For example, following that logic, the label \textit{high} is considered a linguistic value of the variable \textit{Student Performance}. It plays almost the same role as a number but needs to be more precise.
The collection of all linguistic values of a linguistic variable is referred to as a  \textit{term set}.

\begin{table}[h]
\centering
\resizebox{\columnwidth}{!}{%
\begin{tabular}{|l|l|l|}
\hline
\multicolumn{1}{|c|}{\textbf{clean code}} & \multicolumn{1}{c|}{\textbf{functionality}} & \multicolumn{1}{c|}{\textbf{use of inheritance}} \\ \hline
lines of code                   & use of collections                  & use of overriding/overloading \\ \hline
number of fields                & use of own interfaces / build-in    & inherited classes             \\ \hline
number of parameters in methods & number of own exceptions / build-in &                               \\ \hline
use of comments                 & use of comparators                  &                               \\ \hline
number of classes               & use of streams                      &                               \\ \hline
number of methods per class     &                                     &                               \\ \hline
use of serialization            &                                     &                               \\ \hline
\end{tabular}%
}
\caption{The table demonstrates parameters to evaluate code of software projects }
\label{tab:my-table}
\end{table}

\subsubsection{Fuzzy Hedges}
The are two families of modifiers, or hegdes - reinforcing and weakening modifiers. 

The hedge "very" represents the reinforcing modifier:
\begin{equation}
  t_{very}(u) = u^{2}
     \label{veryeq}
\end{equation}

The second family of modifiers is weakening modifiers. For instance, "more-or-less" hedge:
\begin{equation}
  t_{more\mbox{-}or\mbox{-}less}(u) = \sqrt{u}
     \label{mleq}
\end{equation}
Furthermore, "not" hedge is represented as:

\begin{equation}
  t_{not}(u) = 1-u
     \label{mleq}
\end{equation}

Hedges can be applied several times. For example, \textit{not very good performance} is the example of a combined hedge consisting of two atomic hedges \textit{not} and \textit{very}.
\subsubsection{Fuzzy Operations}
The $\alpha $-cut (Alpha cut) is a crisp set that includes all the members 
of the given fuzzy subset f whose values are not less than $\alpha $ for 0< 
$\alpha  \quad  \le $ 1:
\[
f_{\alpha}  = \{x:\mu _{f} (x) \ge \alpha \}
\]
To connect $\alpha $-cuts and 
set operations (A and B are fuzzy sets):

\[
(A \cup B)_{\alpha}  = A_{\alpha}  \cup B_{\alpha}  ,
\quad
(A \cap B)_{\alpha}  = A_{\alpha}  \cap B_{\alpha}  
\]

\subsubsection{Fuzzy Rules}
Fuzzy rules control the output variable. A fuzzy rule is a usual if-then rule containing a condition and conclusion. It has the following form: For example, Rule 15: \textit{If Clean code is Low AND Functionality level is High AND Use of inheritance is Medium THEN Project success is Good.}



\begin{figure}[H]
\centerline{\includegraphics[width=1.0\columnwidth]{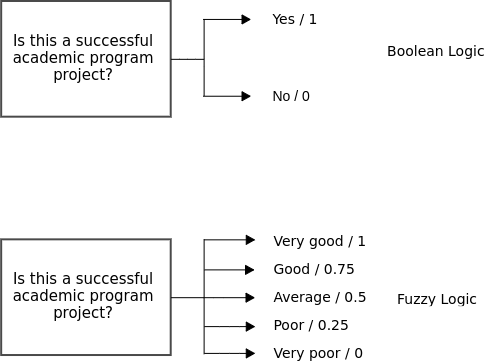}}
\caption{Academic performance evaluation using Classical and Fuzzy sets. }
\label{fig}
\end{figure}

Fuzzy sets have advantages over classical sets when dealing with complex, uncertain, or subjective data (see Fig. \ref{fig}). However, they also have some limitations, such as difficulty defining membership functions and the lack of clear criteria for set membership. 
In traditional grading systems, grades are given based on a fixed set of criteria, such as running a project with no errors and solution independence. Instead of giving a student a letter grade based on a fixed percentage, the fuzzy sets approach can give a grade based on how well the student's performance meets specific criteria.  The instructor must first define the assessment criteria to use fuzzy sets to assess student performance. These criteria can be defined using linguistic variables such as "good," "average," and "poor." 

To apply these rules to a particular student's work, it is necessary to determine the extent to which the project falls into each category. It can be done through various methods, such as self-assessment through questioning, teacher evaluation, or assessment by specific software analysis tool. 




\subsection{Data Collection}
\begin{figure*}[h]
  \begin{subfigure}{0.48\textwidth}
    \includegraphics[width=\linewidth]{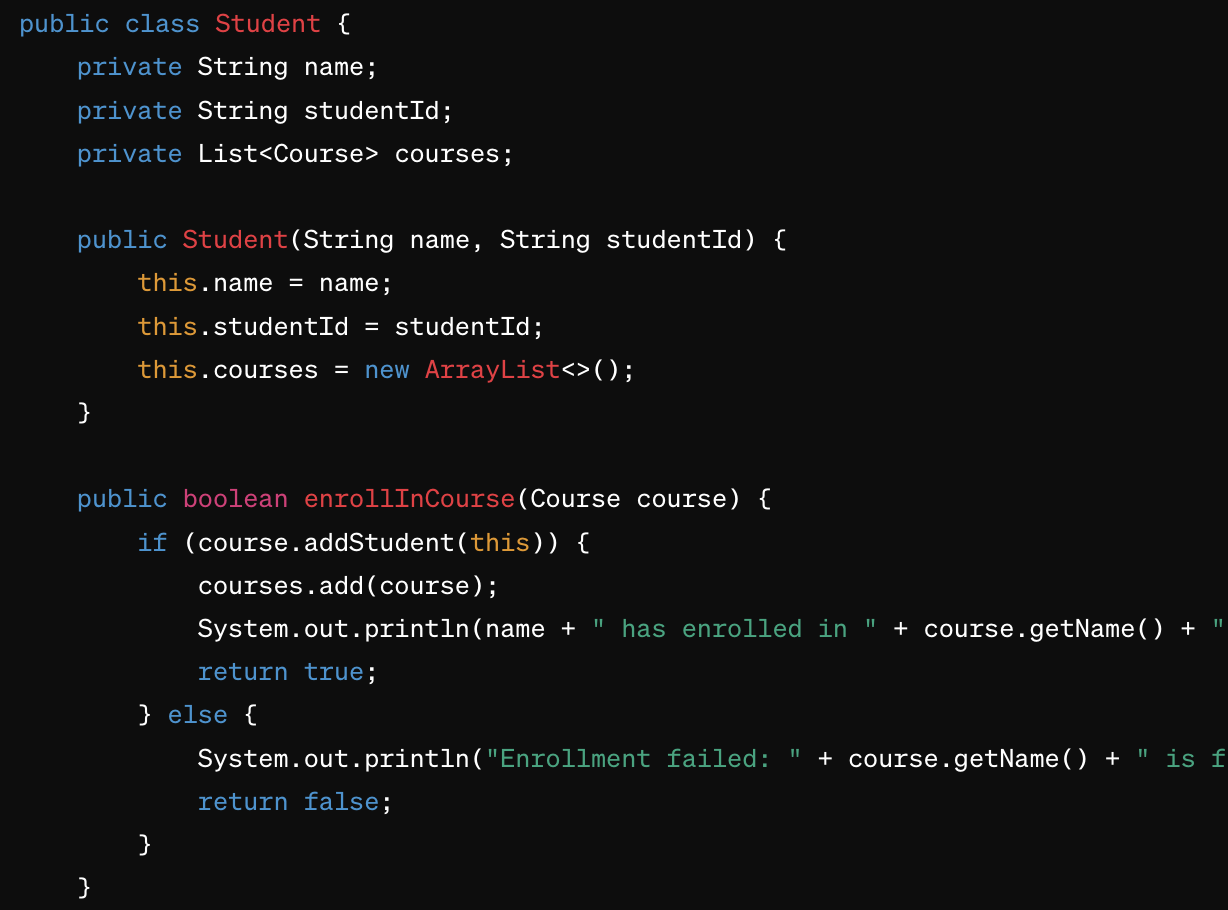}
    \caption{Example of clean code} \label{fig:1a}
  \end{subfigure}%
  \hspace*{\fill}   
  \begin{subfigure}{0.48\textwidth}
    \includegraphics[width=\linewidth]{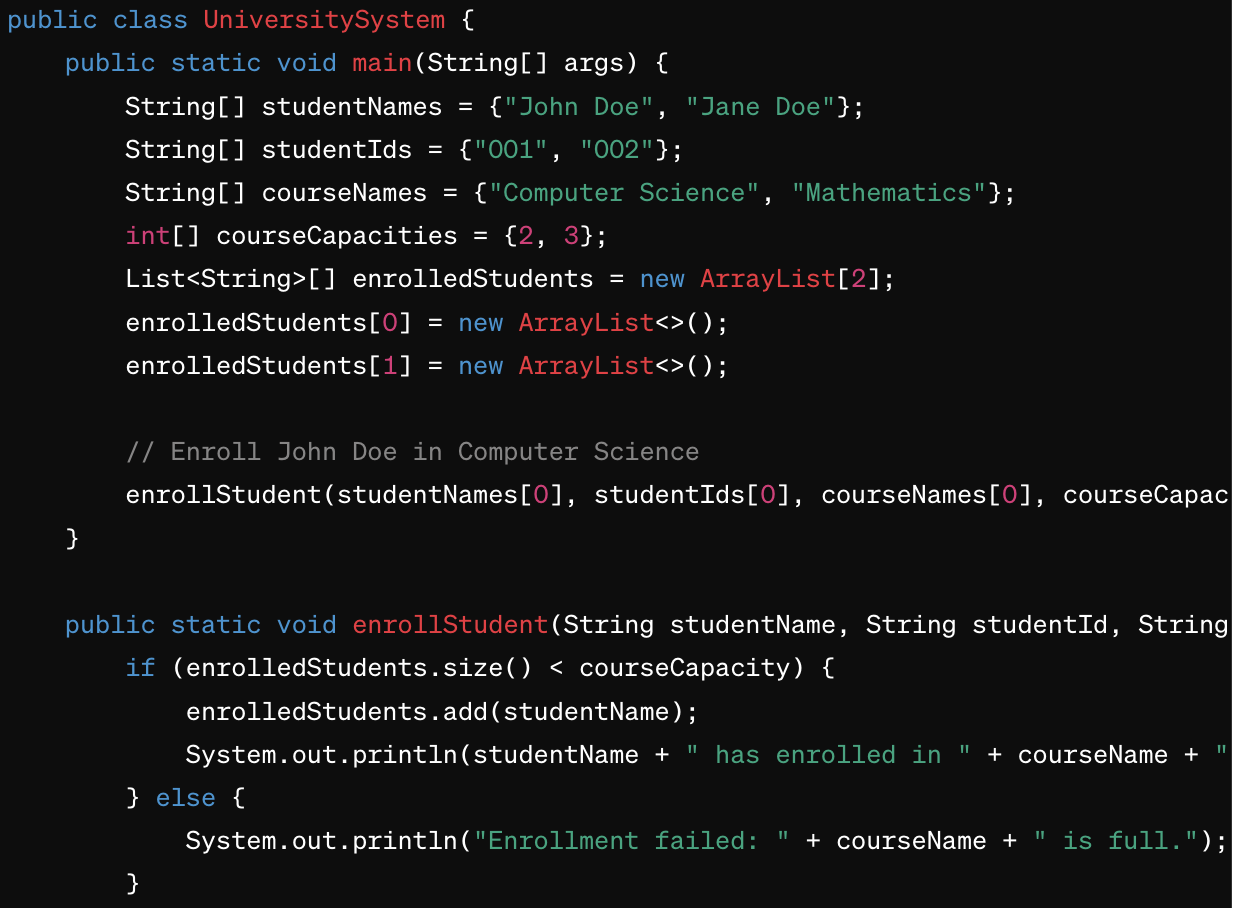}
    \caption{Example of poorly written code} \label{fig:1b}
  \end{subfigure}%
  \hspace*{\fill}   

\caption{Sample project code from the dataset of projects.} \label{fig_code}
\end{figure*}

In this paper, we used two datasets:
\begin{itemize}
    \item \textbf{Projects of students.} We used source codes of 64 projects done in teams by 2nd-year Kazakh-British Technical University students of Information Systems major. Each project was implemented by a team consisting of 4 people.  Fig. \ref{fig_code} presents some code samples from the collected dataset.
    \item \textbf{Students survey data.} We obtained data from a survey (discussed later in the subsection Survey) of students who had completed the Object-Oriented Programming and Design course and three-course instructors. The goal was to identify the key performance indicators required for evaluating OOP projects. The survey was completed by 32 teams, each consisting of four 2-year SITE (school of information technology and engineering) students majoring in information systems. The survey included 21 project-related questions. Fig. \ref{numcl} shows the distribution of the number of classes in the project, and Fig. \ref{marks} shows the distribution of final marks students got for the project.
\end{itemize}

\begin{figure}[h]
\includegraphics[width=0.5\textwidth]{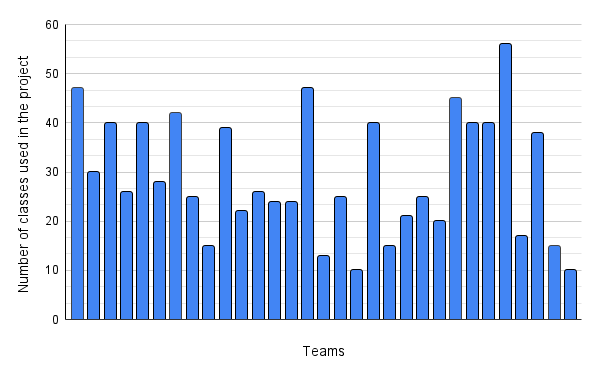}
  \caption{Distribution of the number of classes used in the project}
  \label{numcl}
\end{figure}

\begin{figure}[h]
\includegraphics[width=0.5\textwidth]{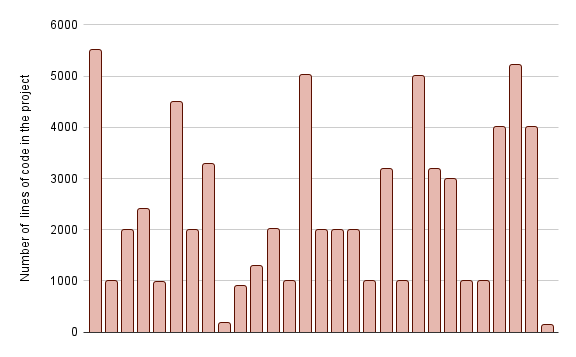}
  \caption{Distribution of the number of lines of code used in the project}
  \label{numcl}
\end{figure}

\begin{figure}[h]
\includegraphics[width=0.5\textwidth]{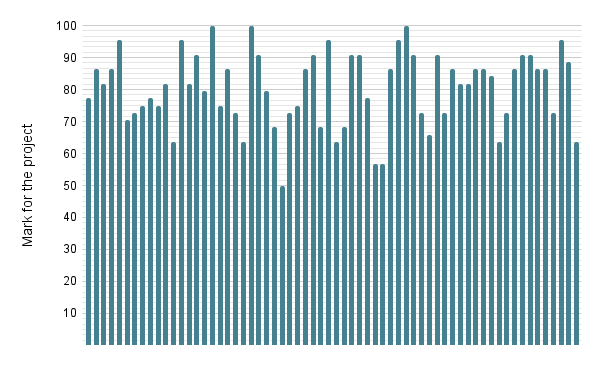}
  \caption{Distribution of the final marks for the project. }
  \label{marks}
\end{figure}

\begin{figure}[h]
\includegraphics[width=0.5\textwidth]{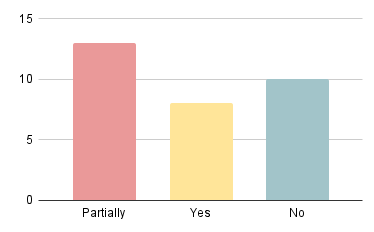}
  \caption{Distribution of equal contribution of team members to the project}
  \label{marks}
\end{figure}

\begin{figure}[h]
\includegraphics[width=0.5\textwidth]{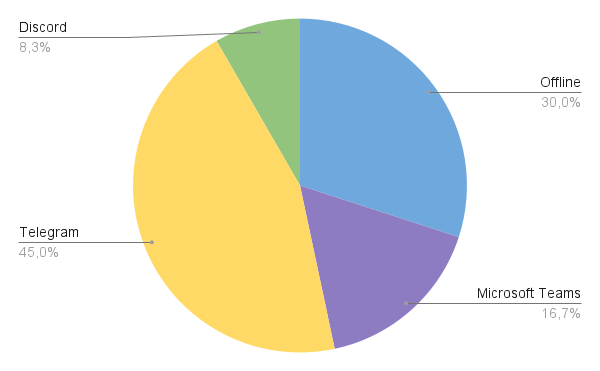}
  \caption{Distribution of frequent ways to communicate with team while working on the project}
  \label{pie}
\end{figure}

\subsection{Survey}
\begin{figure}[H]
\centerline{\includegraphics[width=1.0\columnwidth]{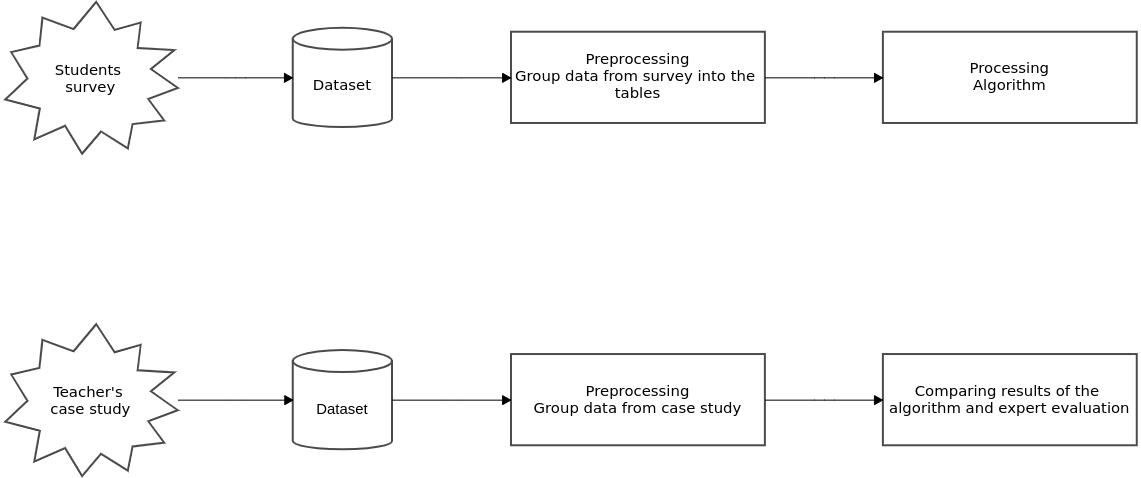}}
\caption{Development of evaluation criteria. Authors use a survey and case analysis with the instructor, and a data set will be compiled to gather information about the criteria for analysing software projects. The selected criteria will be used to analyse the success of the projects.}
\label{fig_sur}
\end{figure}

Various methods can be employed to identify the key fuzzy variables and their corresponding sets, such as surveys, direct rating methods, or consulting experts. We used a data-driven approach; by analyzing the distribution of data points (the mean, the median, the standard deviation), we decided on the parameters of membership functions, considering expert opinions as well. It is a common practice to identify key success factors for a project from a survey (see Fig.\ref{fig_sur}) \cite{agarwal2006defining}. So, in our case, we engaged three experts and conducted a survey among students who had completed the Object-Oriented Programming and Design course. The objective was to pinpoint the key performance indicators crucial for evaluating OOP projects. Working collaboratively with these experts, we identified the necessary fuzzy variables, sets, and partitions. 32 teams, each consisting of four 2-year SITE (school of information technology and engineering) students, participated in the survey. The survey contained 21 project-related questions.

The questionnaire was designed by course experts. Some of the questions were adapted from the book \cite{lethbridge2001object}. The survey contained the following questions:
\begin{itemize}
    \item Team Leader's Name: First and last name of the team leader
    \item Number of Classes Used: Number of classes used in the project (e.g., 37) 
 \item Number of Meetings: Number of online/offline meetings held during the project (e.g., 6-10).
 \item Communication Method: Main methods of communication used with the team (e.g., Offline, Telegram, Discord, etc.)
 \item Equal Contribution: Student's opinion on whether all team members contributed equally (yes/no)
 \item Did you consult with the lecturer or assistants during the project? (yes/no).
 \item How many lines of code are in the project? (e.g., 2000)
 \item How many uncommented lines of code in the project? (e.g., 1800)
 \item How many methods per class on average in the project? (e.g., 10)
 \item How many public methods per class on average in the project? (e.g., 3)
 \item How many public instance variables per class on average in the project? (e.g., 4)
 \item How many parameters per method on average in the project? (e.g., 3)
 \item How many lines of code per method on average in the project? (e.g., 30)
 \item Choose an appropriate distribution of tasks during the project (e.g., The leader took most of the work himself, gave minor tasks to the team)
 \item On a scale of 0 to 10, rate the team leader's performance during the project
 \item If you used patterns, what patterns did you use to design the project? (e.g., decorator)
\item Documentation Creation: Whether certain types of documentation were created (e.g., software requirements specification document).
\item Group Communication: Whether a group was created for communication.
\item Importance to Career: How important the student thinks the project is to their future career.
\end{itemize}
As a result, we obtained a dataset containing information about a final Object-Oriented Programming (OOP) project evaluation, with each row representing a student's responses. We also extended the dataset with the real marks students obtained for their project. 
Table \ref{X} shows the statistics of certain project features explored in the survey.

\begin{table}[h]
\begin{tabular}{|l|l|l|l|}
\hline
\textbf{Feature          }                  & \textbf{Average} & \textbf{Maximum  }    & \textbf{Minimum }\\ \hline
Number of classes                  & 29.2    & 56           & 10      \\ \hline
Number of meetings                 & -       & More than 15 & 0-5     \\ \hline
Number of methods per class        & -       & 50-60        & 3       \\ \hline
Number of lines of code            & 2406.2  & 5500         & 130     \\ \hline
Number of lines of code per method & -       & 90           & 3-7     \\ \hline
Mark                               & 82.5    & 100          & 56.8    \\ \hline
\end{tabular}
\label{X}
\caption{Academic software project characteristics (data from students survey)}
\end{table}

\begin{figure}[h]
\includegraphics[width=0.5\textwidth]{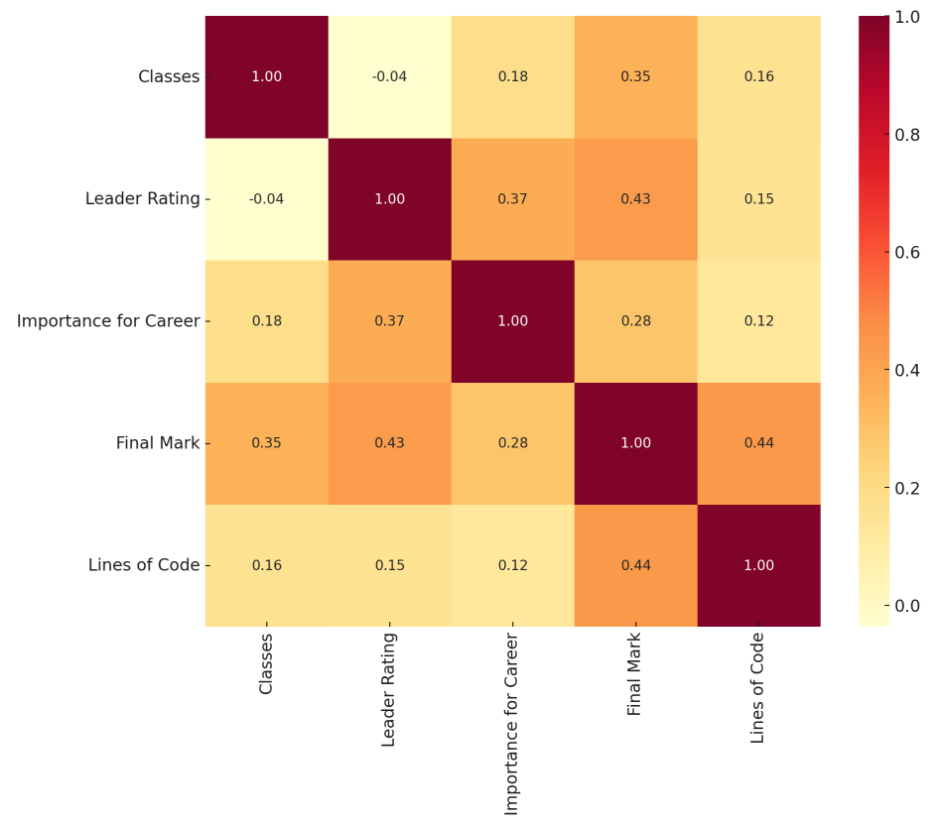}
  \caption{Correlation heatmap of project evaluation metrics}
  \label{heatmap}
\end{figure}
\begin{figure*}[h]
  \includegraphics[width=\textwidth]{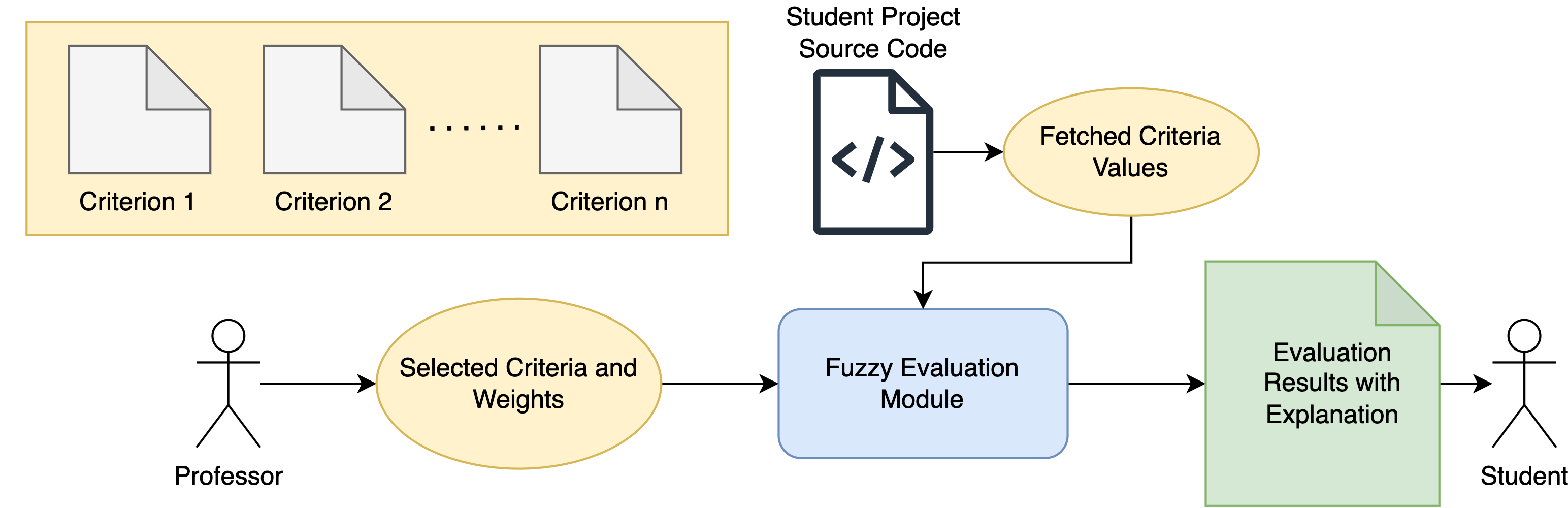}
  \caption{The Proposed Evaluation Methodology.}
  \label{archit}
\end{figure*}

\begin{figure*}[h]
  \begin{subfigure}{0.5\textwidth}
    \includegraphics[width=\linewidth]{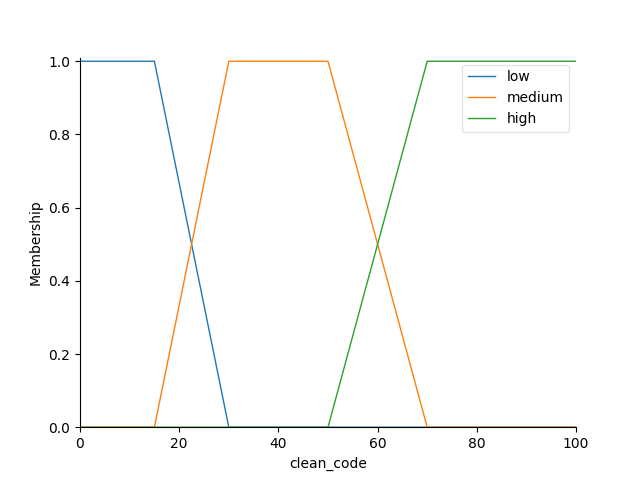}
  \end{subfigure}%
  \begin{subfigure}{0.5\textwidth}
    \includegraphics[width=\linewidth]{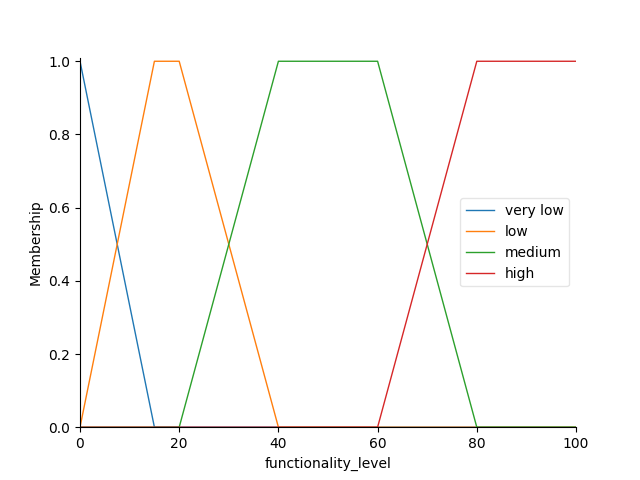}
  \end{subfigure}%
  \\
    \begin{subfigure}{0.5\textwidth}
    \includegraphics[width=\linewidth]{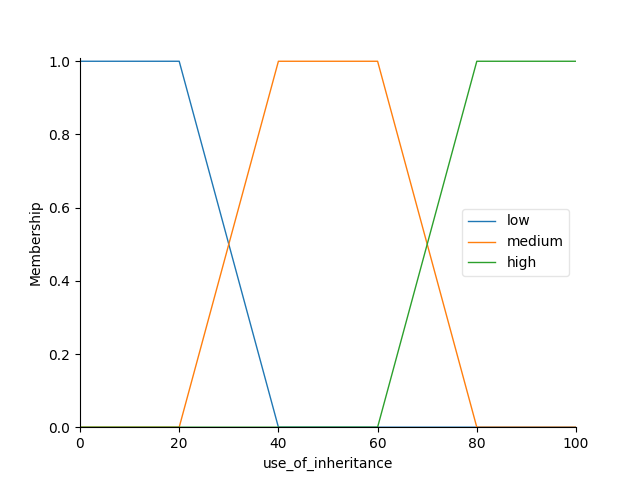}
  \end{subfigure}%
  \begin{subfigure}{0.5\textwidth}
    \includegraphics[width=\linewidth]{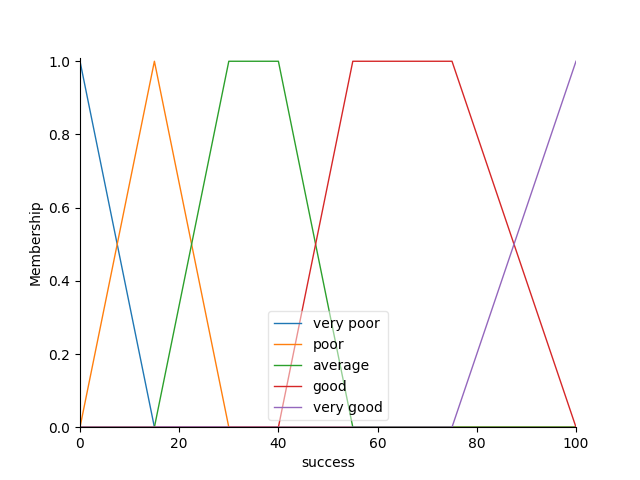}
  \end{subfigure}%
 \caption{Input fuzzy sets for \textit{Clean code }, \textit{Functionality level},\textit{ Use of inheritance} and Output fuzzy sets for \textit{Success }} 
\label{fig:fuzzy_sets}
\end{figure*}

Fig. \ref{heatmap} presents the correlation heatmap of project evaluation metrics built using numerical data extracted from a survey. The heatmap palette ranges from light yellow (indicating lower correlation) to deep orange (indicating higher correlation). 
The following observations can be done:
\begin{itemize}
    \item \textbf{Final Mark and Classes.} Moderate positive correlation (0.35). So, having more classes in the project might be associated with slightly higher final marks, potentially reflecting a more complex project structure with bigger functionality.
\item \textbf{Final Mark and Leader Rating.} Moderate positive correlation (0.43).  Effective leadership likely contributes to better project outcomes.
\item \textbf{Final Mark and Lines of Code.} A moderate positive correlation (0.44) suggests that projects with more lines of code tend to receive higher marks. This might indicate that larger or more complex projects, which require more code, are viewed favorably in evaluations, assuming the quality of the code is also high.
\end{itemize}

These correlations reveal how various factors related to project management and execution can influence the overall evaluation of a project.



\subsection{Proposed Methodology}
The proposed intelligent system for evaluating academic software projects using a fuzzy inference system. is presented in Fig. \ref{archit}.

Table \ref{terms} shows the information about term sets of the input and output variables and their domains. Fig. \ref{fig:fuzzy_sets} presents the membership functions for all fuzzy variables.

We partition the spectrum of possible assessments corresponding to linguistic tags \cite{jaciii}. We have three input variables describing the project - \textit{Clean Code}, \textit{Functionality Level}, \textit{Use of Inheritance}.  The output variable is \textit{Project Success} (see Fig. \ref{fig:fuzzy_sets}). As can be seen, we have \textit{'Low'}, \textit{'Medium'}, and\textit{ 'High'} fuzzy sets for \textit{Clean Code} and \textit{Use of Inheritance} input variables, \textit{'Very Low', 'Low', 'Medium',} and \textit{'High' }fuzzy sets for \textit{Functionality Level} and \textit{'Very Poor', 'Poor', 'Average', 'Good',} and \textit{'Very Good'} for the output variable. The linguistic expressions for the fuzzy model's output variable were partly adapted from \cite{output}.

 We use fuzzy rules to build fuzzy relationships between input and output variables. Our fuzzy inference system has 36 fuzzy rules, as shown in Table \ref{rules}. Fuzzy if-then rules are descriptive and usually created by experts or human knowledge.


\begin{table}[h]
\caption{Fuzzy attributes of the fuzzy inference system.}
\begin{tabular}{|l|l|l|}
\hline
\textbf{Fuzzy Variable} & \textbf{Term Set} & \textbf{Domain} \\ \hline
Clean Code                 &      T = \{Low, Medium, High\}                    &          X=[0,100]                      \\ \hline
Functionality Level               &    T = \{Very Low, Low, Medium, High\}                     &       X=[0,100]                                 \\ \hline
Use of Inheritance &           T = \{Low, Medium, High\}                 &                                 X=[0,100]       \\ \hline
                Project Success     &       T = {Very Poor, Poor, Average, Good, Very Good}                &                       X=[0,100]            &               \hline
                    
\end{tabular}
\label{terms}
\end{table}

Based on the survey responses and instructors' opinions, fuzzy rules were created to determine the relationship between the evaluation criteria and the output linguistic variables. We identified these fuzzy evaluation criteria and fuzzy rules by working collaboratively with one professor and two instructors, paying attention to survey results. These rules form the basis for evaluating academic software projects (see Table \ref{rules}).

\begin{table}[tb]
\caption{Fuzzy rules.}
\begin{tabular}{|p{1cm}|p{1cm}|p{1.5cm}|p{1.5cm}|p{1cm}|}
\hline
\multicolumn{1}{|l|}{\textbf{rule}} & \textbf{clean code} & \textbf{functionality level} & \textbf{use of inheritance} & \textbf{project success} \\ \hline
\cellcolor 1           & High                 & High                          & High                          & Very Good        \\ \hline
2                                   & Medium               & High                          & High                          & Very Good        \\ \hline
3                                   & Low                  & High                          & High                          & Good             \\ \hline
4                                   & High                 & Medium                        & Medium                        & Good             \\ \hline
5                                   & Medium               & Medium                        & Medium                        & Average          \\ \hline
6                                   & Low                  & Medium                        & Medium                        & Average          \\ \hline
7                                   & High                 & Low                           & Low                           & Poor             \\ \hline
8                                   & Medium               & Low                           & Low                           & Very Poor        \\ \hline
9                                   & Low                  & Low                           & Low                           & Very Poor        \\ \hline
10                                  & High                 & Very Low                      & High                          & Average          \\ \hline
11                                  & Medium               & Very Low                      & High                          & Poor             \\ \hline
12                                  & Low                  & Very Low                      & High                          & Poor             \\ \hline
13                                  & High                 & High                          & Medium                        & Very Good        \\ \hline
14                                  & Medium               & High                          & Medium                        & Good             \\ \hline
15                                  & Low                  & High                          & Medium                        & Good             \\ \hline
16                                  & High                 & Medium                        & Low                           & Average          \\ \hline
17                                  & Medium               & Medium                        & Low                           & Average          \\ \hline
18                                  & Low                  & Medium                        & Low                           & Poor             \\ \hline
19                                  & High                 & Low                           & High                          & Average          \\ \hline
20                                  & Medium               & Low                           & High                          & Average          \\ \hline
21                                  & Low                  & Low                           & High                          & Poor             \\ \hline
22                                  & High                 & Very Low                      & Medium                        & Poor             \\ \hline
23                                  & Medium               & Very Low                      & Medium                        & Poor             \\ \hline
24                                  & Low                  & Very Low                      & Medium                        & Very Poor        \\ \hline
25                                  & High                 & High                          & Low                           & Good             \\ \hline
26                                  & Medium               & High                          & Low                           & Average          \\ \hline
27                                  & Low                  & High                          & Low                           & Poor             \\ \hline
28                                  & High                 & Medium                        & High                          & Very Good        \\ \hline
29                                  & Medium               & Medium                        & High                          & Good             \\ \hline
30                                  & Low                  & Medium                        & High                          & Average          \\ \hline
31                                  & High                 & Low                           & Medium                        & Average          \\ \hline
32                                  & Medium               & Low                           & Medium                        & Average          \\ \hline
33                                  & Low                  & Low                           & Medium                        & Poor             \\ \hline
34                                  & High                 & Very Low                      & Low                           & Poor             \\ \hline
35                                  & Medium               & Very Low                      & Low                           & Poor             \\ \hline
36                                  & Low                  & Very Low                      & Low                           & Very Poor        \\ \hline
\end{tabular}
\label{rules}
\end{table}




In a more general case,  when it is considered time-consuming to survey experts and students, collecting criteria and their importance can be done via the system. For example, let $C= \{C_1, C_2,...,C_n\} $ be the evaluation criteria that an instructor or professor chooses, e.g., the code clarity, documentation, etc. Then $w=\{w_1, w_2,... w_n \}$ are the weights (importance) of the corresponding criteria chosen by the course instructor. Next, $P = \{P_1, P_2, ..., P_n\}$ are student projects.  $L = \{ L_1, L_2,...,L_n \}$ represents linguistic variables with the corresponding term set representing its assessment, a vector of linguistic terms on $L_i$  - \textit{\{High, Average, Low\}}.
 
 Then, a student project can be evaluated based on weight-based fuzzy rules generation \cite{frulle}, \cite{frulle2}. If the weight is high, a mark for this criterion is essential. After a case study with professors about their evaluation methods, we concluded that the most essential criterion greatly influences the final result. For example, we have three criteria - \textit{Clean code, The use of Inheritance, Functionality} with respective weights selected by the Professor as, for example, \textit{Medium, Low, High}. The code quality score was obtained by summing up all related scores automatically extracted from the source code, including the number of methods, following naming conventions, etc. 
 
 



\section{Results}
\subsection{Simulation and Performance Evaluation}

\begin{figure*}[h]
  \begin{subfigure}{0.5\textwidth}
    \includegraphics[width=\linewidth]{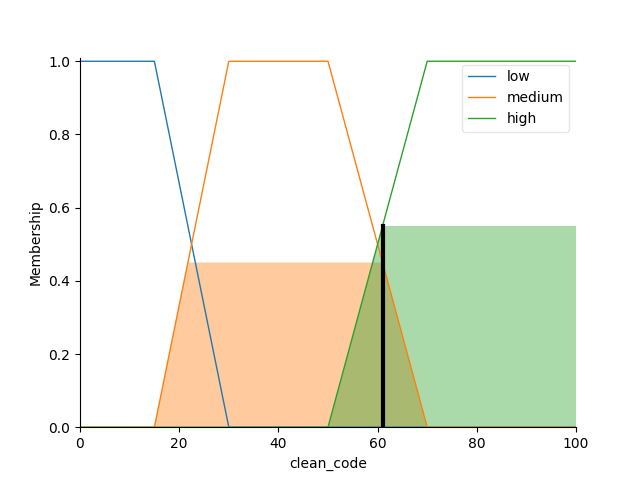}
    \caption{Applying input 61\% on \textit{Clean code} fuzzy set} \label{fig:1a}
  \end{subfigure}%
  \hspace*{\fill}   
  \begin{subfigure}{0.5\textwidth}
    \includegraphics[width=\linewidth]{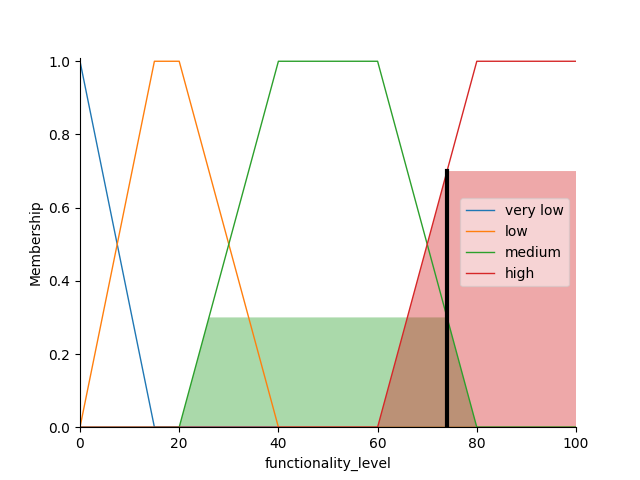}
    \caption{Applying input 74\% on \textit{ Functionality level} fuzzy set} \label{fig:1b}
  \end{subfigure}%
  \hspace*{\fill}   
  \\
  \begin{subfigure}{0.5\textwidth}
    \includegraphics[width=\linewidth]{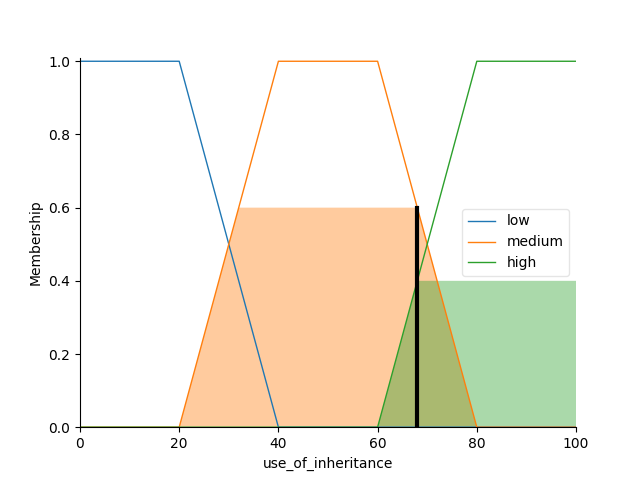}
    \caption{Applying input 68\% on \textit{ Use of inheritance} fuzzy set} \label{fig:1a}
  \end{subfigure}%
  \begin{subfigure}{0.5\textwidth}
    \includegraphics[width=\linewidth]{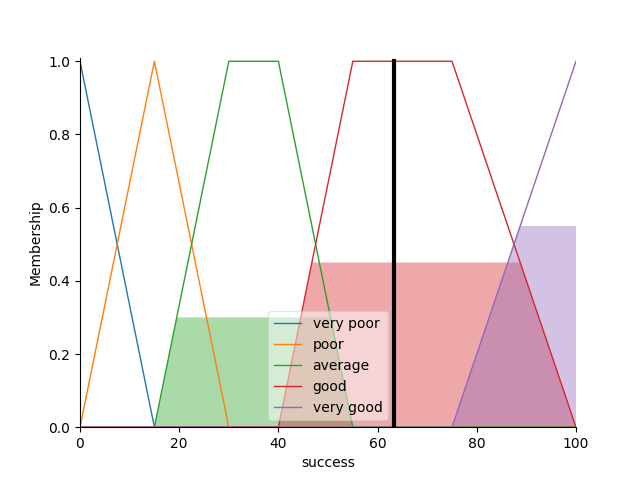}
    \caption{Aggregated Membership and Result, 63.27\%} 
  \end{subfigure}
\caption{Simulation Results.} \label{fig_ex}
\end{figure*}

We can now simulate our fuzzy inference system by specifying the inputs and using the defuzzification approach. For example, let us consider the following input data and determine the overall project success: The \textit{Clean code}, \textit{Functionality}, and \textit{Use of inheritance} are 61\%, 74\%, and 68\% respectively. The output membership functions are then mixed with the maximum operator (fuzzy aggregation). Next, in order to obtain a clear answer, we must do defuzzification, which we accomplish using the centroid approach. Fuzzy rule-based aggregation yields 63.27 \% as the total project success. Fig. \ref{fig_ex} shows the visualized result.

\begin{table}[h]
\begin{tabular}{|l|r|r|r|r|r|}
\hline
          & \multicolumn{1}{l|}{\textbf{Clean Code}} & \multicolumn{1}{l|}{\textbf{Functionality}} & \multicolumn{1}{l|}{\textbf{Use of Inheritance}} & \multicolumn{1}{l|}{\textbf{Proposed method}} & \multicolumn{1}{l|}{\textbf{Real mark}} \\ \hline
Project 1 & 100                                      & 82                                          & 84                                               & 92                                           & 95                                      \\ \hline
Project 2 & 67                                       & 34                                          & 100                                              & 64                                           & 57                                      \\ \hline
Project 3 & 100                                      & 63                                          & 100                                              & 91                                           & 86                                      \\ \hline
\end{tabular}
\label{results}
\end{table}

To assess the effectiveness of the proposed fuzzy intelligence system, three instructors from an Object-Oriented Programming (OOP) course were enlisted. Each instructor evaluated three student projects manually, using predefined evaluation criteria, and the mean of their evaluation was taken. These evaluations were then compared with assessments conducted by the fuzzy intelligent system. The results of this comparative analysis are presented in Table X. The findings indicate that the fuzzy intelligent system performed robustly, demonstrating promising results that aligned with the manual evaluations conducted by the course instructors.




\subsection{Application Prototype}

\begin{figure*}[h]
\centering
  \includegraphics[width=0.7\textwidth]{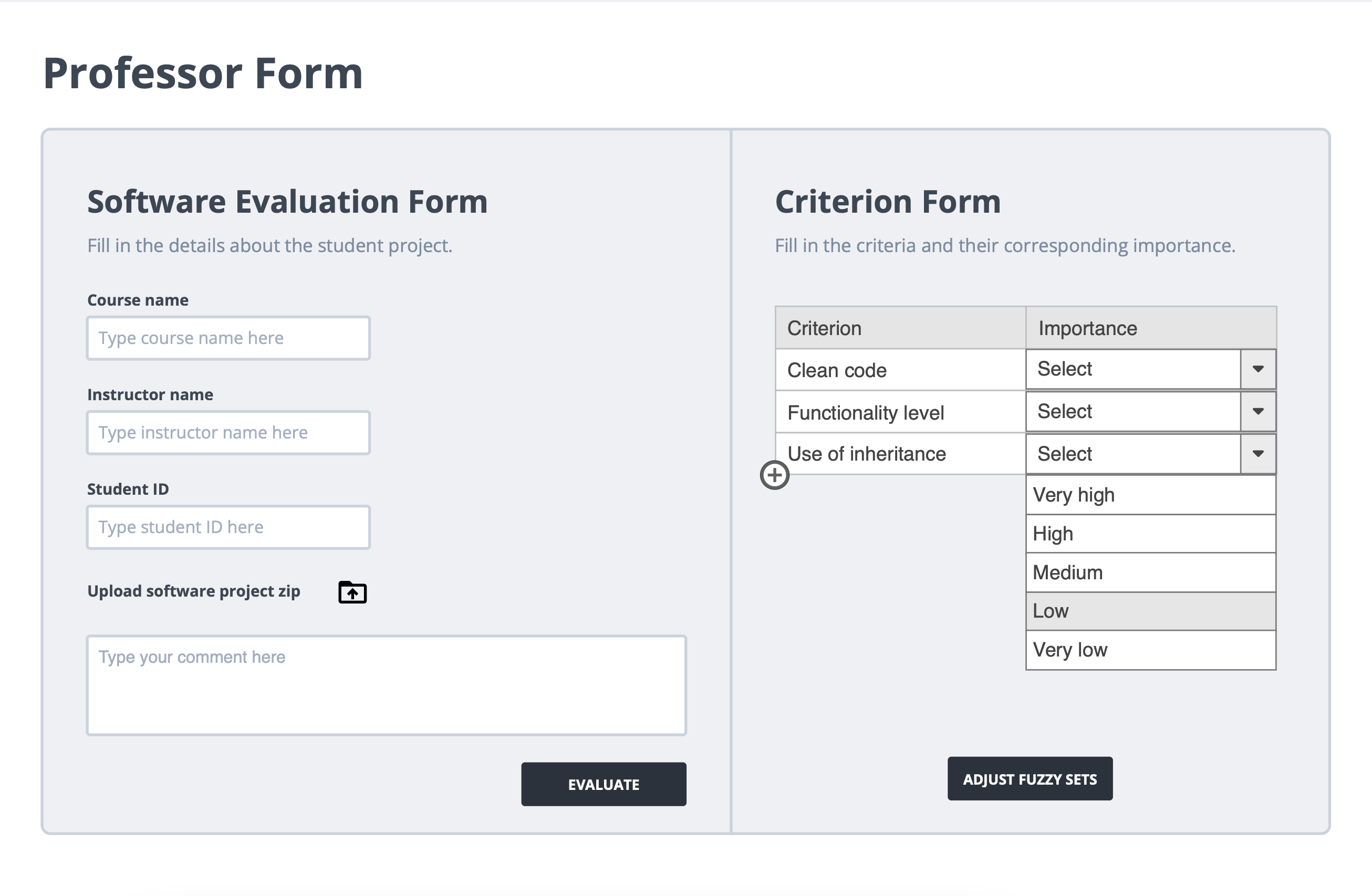}
  \caption{Prototype: Project Evaluation Professor Form.}
  \label{evform}
\end{figure*}
\begin{figure}[h]
  \includegraphics[width=0.4\textwidth]{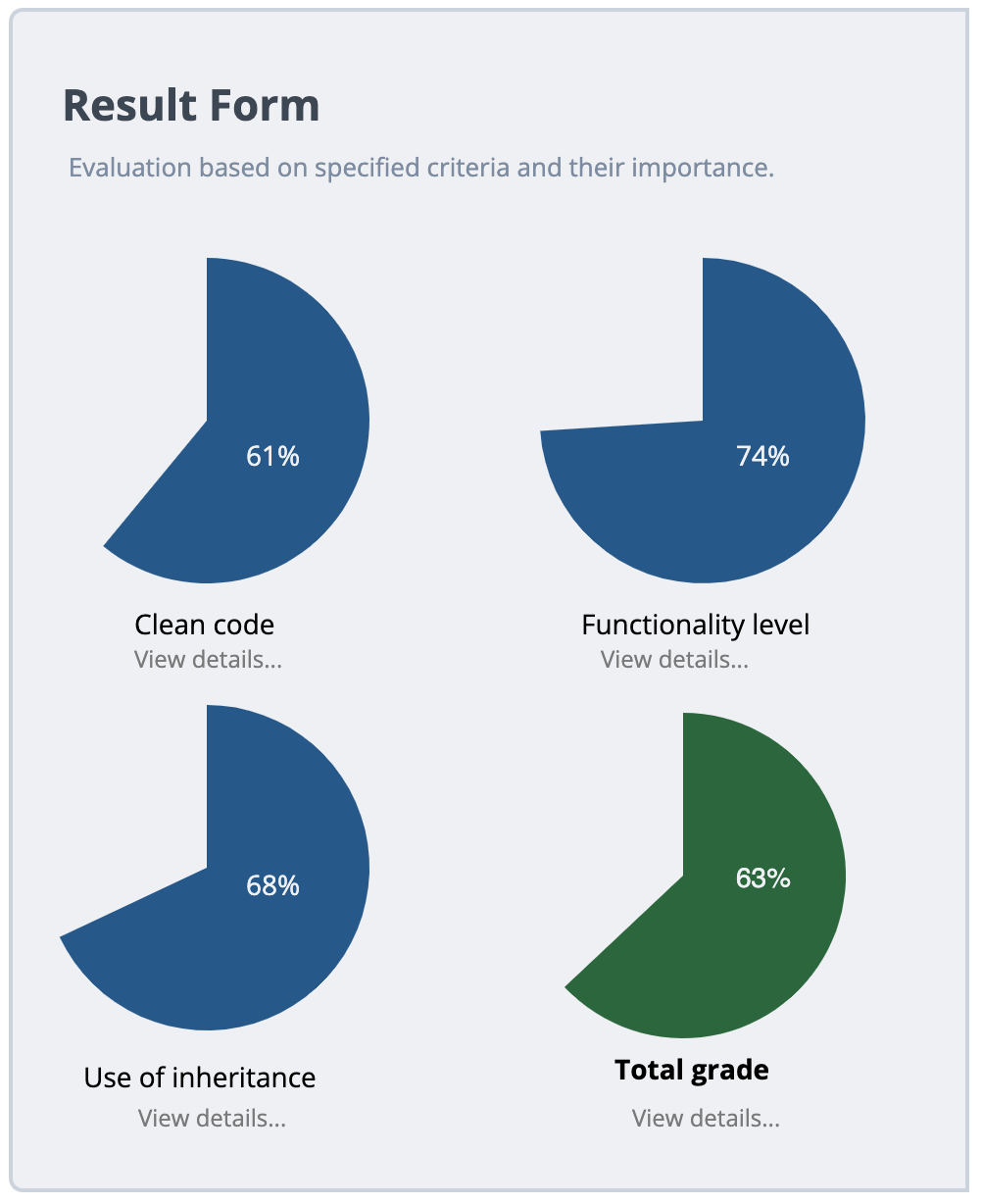}
  \caption{Evaluation Report}
  \label{report}
\end{figure}

Fig. \ref{evform} shows the layout of a professor evaluation form specifically designed for evaluating software projects. The professor has to enter data such as course name, personal information, and student data and then upload the code of the student's software project into the form. After entering the preliminary data about the software project, the instructor can select different evaluation criteria with appropriate weights reflecting the importance of each criterion. Ultimately, the form generates recommendations for evaluating student work based on weighted criteria, simplifying the grading process and ensuring a fair and objective evaluation of the student's project that meets educational standards and expectations.

 Such an intelligent system can be integrated with the SonarLint code analyzer as an alternative to manually evaluating the criteria or taking them from students' surveys. The system includes a user-friendly interface for entering project data and evaluating it based on predefined fuzzy rules. It can analyze project source code and provide evaluation results in an understandable format. Fig. \ref{report} shows the project evaluation final report page prototype.

The proposed system cannot replace the teacher. However, it helps evaluate the project by analyzing the student's work based on the given criteria. The user needs to specify the criteria and their weight for evaluation, indicate the documentation on the project, and upload the project's source code to get the algorithm's result. The system analyses the number of classes in the project, the average length of code in classes, the number of fields and methods, and the use of access modifiers (private/package/public). Also, with the help of the connected anti-plagiarism service, the system checks the project's uniqueness based on the uploaded works and information on the Internet and gives the plagiarism percentage.

\section{Conclusion}
This paper proposes a novel approach for evaluating software projects in an academic environment. By implementing a fuzzy intelligent system, we aim to automate the evaluation process, reduce subjective biases, and manage the increasing instructor workload effectively. We surveyed students and faculty, and the responses helped to identify key evaluation factors in evaluating academic software projects. The fuzzy system uses predefined criteria - clean code, use of inheritance, and functionality - transformed into fuzzy sets and employs a fuzzy inference mechanism defined in collaboration with educational experts.



Fuzzy set theory served as the basis for our evaluation model, allowing us to represent imprecise and subjective information related to project evaluation.

The study results can help academic instructors save time, reduce costs, and improve the quality and efficiency of evaluating student software projects. In turn, students can get faster feedback on their work and analyze the code of their software projects.

As for the limitations, the system may not easily adapt to course content or evaluation standards changes without significantly reconfiguring the fuzzy sets and rules. Another limitation is subjectivity in the criteria definition. The fuzzy sets partitions depend on experts' subjective judgments. So, the possible improvement for future works can involve a more objective method for defining evaluation criteria using data analytics and machine learning to analyze historical project data. Another improvement we plan is the incorporation of teamwork as an evaluation parameter.



\bibliography{export}

\end{document}